\documentclass[aps,prl,twocolumn,superscriptaddress,showpacs]{revtex4}
\usepackage{graphicx}
\usepackage{mathrsfs}
\usepackage{bm}

\begin{document}

\title{Quantum Anomalous Hall Effect in Graphene from Rashba and Exchange Effects}

\author{Zhenhua Qiao}
\affiliation{Department of Physics, The University of Texas, Austin,
78712, USA}
\author{Shengyuan A. Yang}
\affiliation{Department of Physics, The University of Texas, Austin,
78712, USA}
\author{Wanxiang Feng}
\affiliation{Beijing National Laboratory for Condensed Matter
Physics, and Institute of Physics, Chinese Academy of Sciences,
Beijing 100190, China}
\author{Wang-Kong Tse}
\affiliation{Department of Physics, The University of Texas, Austin,
78712, USA}
\author{Jun Ding}
\affiliation{Beijing National Laboratory for Condensed Matter
Physics, and Institute of Physics, Chinese Academy of Sciences,
Beijing 100190, China}
\author{Yugui Yao$^*$}
\affiliation{Beijing National Laboratory for Condensed Matter
Physics, and Institute of Physics, Chinese Academy of Sciences,
Beijing 100190, China} \affiliation{Department of Physics, The
University of Texas, Austin, 78712, USA}
\author{Jian Wang}
\affiliation{Department of Physics and the Center of Theoretical and
Computational Physics, The University of Hong Kong, Hong Kong, China
}
\author{Qian Niu}
\affiliation{Department of Physics, The University of Texas, Austin,
78712, USA}

\begin{abstract}
We investigate the possibility of realizing quantum anomalous Hall
effect in graphene. We show that a bulk energy gap can be opened in
the presence of both Rashba spin-orbit coupling and an exchange
field. We calculate the Berry curvature distribution and find a
non-zero Chern number for the valence bands and demonstrate the
existence of gapless edge states. Inspired by this finding, we also
study, by first principles method, a concrete example of graphene
with Fe atoms adsorbed on top, obtaining the same result.
\end{abstract}
\maketitle

Graphene is a fascinating material not only because of its
extraordinary electrical, magnetic, and mechanical properties, but
also because of its Dirac-cone structure in its low energy spectrum,
which involves three binary internal degrees of freedom: spin,
sublattice and
valley~\cite{ExpGraphene1,ExpGraphene2,RMPGraphene1,RMPGraphene2}.
Manipulating these degrees of freedom may turn out to be the most
rewarding means of controlling the properties of graphene and
related materials. For example, by breaking the AB sublattice
symmetry, an energy gap can be opened at the Dirac
points~\cite{Stagger1,Stagger2,Stagger3,Stagger4}, leading to the
valley-contrasting Hall, magnetic and optical
responses~\cite{XiaoDi}. The intrinsic spin-orbit coupling~(SOC) is
proposed to give rise to the quantum spin Hall
effect~\cite{Kane1,Kane2}, a quantized response of a transverse spin
current to an electric field.

However, the intrinsic SOC in graphene is proved to be too weak to
realize the quantum spin Hall effect under present experimental
conditions~\cite{WeakSOI1,WeakSOI2}. On the other hand, the
extrinsic Rashba SOC from breaking the mirror symmetry of the
graphene plane tends to destroy this effect~\cite{Kane1}. Rashba SOC
can be very large for graphene grown on a
substrate~\cite{SOINi1,SOINi2}. For example, a 225~\textrm{meV} of
Rashba spin splitting has been observed for graphene grown on
Ni~\cite{SOINi2}. We expect that Rashba SOC should also be present
for graphene on insulating substrate, or with dilute adsorbates. In
the following, we show that although Rashba SOC is detrimental to
the quantum spin Hall effect, it helps to realize another important
topological phenomenon: the quantum anomalous Hall effect.

This effect is signaled by a quantized charge Hall conductance for
an insulating state. Unlike the quantum spin Hall effect, the
quantum anomalous Hall effect requires the breaking of time reversal
symmetry. Unlike the quantum Hall effect, which is induced by a
strong magnetic field, the quantum anomalous Hall effect relies on
the internal magnetization and SOC. The quantized value of the Hall
conductance is related to a bulk topological number and is robust
against disorder and other perturbations~\cite{Thouless,Kohmoto}.
Although there have been many theoretical studies of this unusual
effect~\cite{haldane,SCZhang1,Nagaosa1,SCZhang3,Congjun}, no
experimental observation has been reported so far.

In this Letter, we predict that the quantum anomalous Hall effect
can be realized in graphene by introducing Rashba SOC and an
exchange field. We also consider a concrete example with transition
metal atoms (\emph{e.g.,} Fe) adsorbed on top of graphene. The
resulting broken structural symmetry gives rise to a Rashba type SOC
whereas the hybridization between the carbon $\pi$ state and the
3\emph{d}-shell states of magnetic atoms produces a macroscopic
exchange field. The Hall conductivity is found to be
$\sigma_{yx}=2e^2/h$ when the Fermi level lies in the bulk gap. From
first principles calculations, we find that a bulk gap $\sim
5.5~\textrm{meV}$ can be opened by adsorbing Fe atoms on top of
graphene, which is readily accessible under current experimental
conditions.

In the tight-binding approximation, the Hamiltonian for graphene in
the presence of Rashba SOC and exchange field can be written as:

\begin{eqnarray}
H=&-&t \sum_{\langle{ij}\rangle \alpha }{ c^\dagger_{i
\alpha}c_{j\alpha}+ {i} t_{so}\sum_{\langle{ij}\rangle \alpha \beta
}\hat{\mathbf{e}}_{z}{\cdot}(\bm{\sigma_{\alpha
\beta}}{\times}{\mathbf{d}}_{ij})c^\dagger_{i \alpha } c_{j \beta }} \nonumber \\
&+& \lambda\sum_{i\alpha}{ c^\dagger_{i\alpha}\sigma_{z}c_{i\alpha}}
\label{eq1}
\end{eqnarray}
where $c^\dag_{i}$($c_{i}$) is the electron creation (annihilation)
operator on site $i$, and $\bm{\sigma}$ are the Pauli matrices. The
angular bracket in $\langle i,j\rangle$ stands for nearest
neighboring sites. The first term is the usual nearest neighbor
hopping term. The second term is the Rashba SOC with coupling
strength $t_{so}$, and ${\mathbf{d}}_{ij}$ represents a unit vector
pointing from site $j$ to site $i$. The third term corresponds to a
uniform exchange field~\cite{Magnetism}. Throughout this Letter, we
measure the Fermi level, Rashba SOC, and exchange energy in units of
the hopping parameter $t$.

\begin{figure}
\includegraphics[width=7cm,totalheight=6.cm,angle=0]{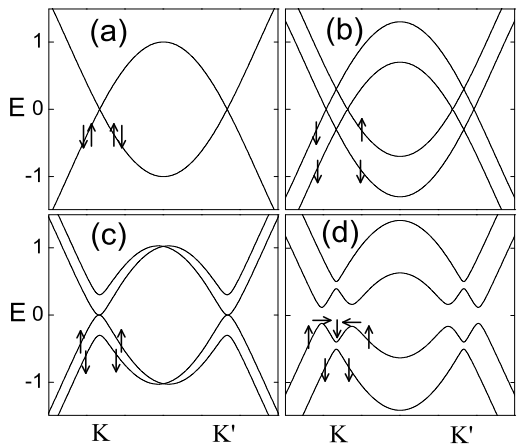}
\caption{Evolution of band structures of bulk graphene along the
profile of $k_y=0$. Arrows represent the spin directions. (a)
Pristine graphene: spin-up and spin-down states are degenerate; (b)
When only exchange field $\lambda=0.4$ is applied, the
spin-up/spin-down bands are upward/downward lifted with the four
bands crossing near $K$ and $K'$ points; (c) When only Rashba
SOC $t_{so}=0.1$ is present, the spin-up and spin-down states are
mixed around the band crossing points; (d) When both exchange field
$\lambda=0.4$ and Rashba SOC $t_{so}=0.1$ are present, a bulk gap
is opened and all four bands become non-degenerate.} \label{fig1}
\end{figure}

By transforming the Hamiltonian of Eq.(\ref{eq1}) into a $4 \times
4$ matrix $H(\bm{k})$ for each crystal momentum, the band structure
of bulk graphene can be numerically obtained by diagonalizing
$H(\bm{k})$. Fig.\ref{fig1} shows the evolution of the band
structures with Rashba SOC $t_{so}$ and exchange field $\lambda$.
Panel (a) plots the band structure of the pristine graphene with
Dirac cones centered at $K$ and $K'$ points in the reciprocal space.
Due to spin degeneracy, $K$ and $K'$ points are four-fold
degenerate, while other points are doubly-degenerate. When only the
exchange field is applied, the spin-up (spin-down) bands are pushed
upward (downward) as shown in panel (b). When only Rashba SOC is turned on,
spin-up and spin-down states are mixed around the band crossing
points and spin degeneracy is lifted. A bulk gap is opened when both
Rashba SOC and exchange field are present, and the four bands become
completely non-degenerate. The presence of such a bulk gap indicates an
insulating state. As we argue below, this insulating state is
topologically nontrivial with gapless chiral edge states and
exhibiting a quantized charge Hall conductance.

\begin{figure}
\includegraphics[width=7.cm,totalheight=9cm,angle=0]{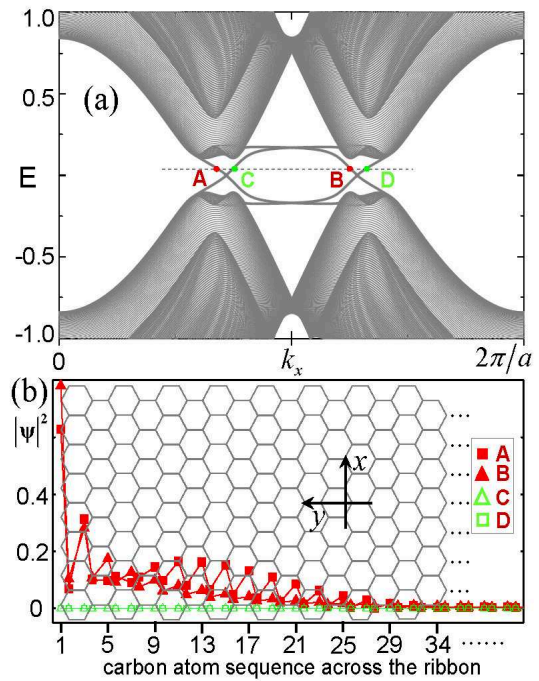}
\caption{(Color online) (a) Energy spectrum of zigzag-edged graphene
ribbons with $t_{so}=0.2$ and $\lambda =0.18$. The Fermi level
$E=0.06$ corresponds to four different edge states $A$, $B$, $C$,
and $D$. $a$ is the lattice constant. (b) Wave function
distributions $|\psi|^2$ across the width for the four edge states
(only part of the ribbons are shown): $A,B$ states are localized at
the left boundary, with noticeable values of $|\psi|^2$ on about 30
carbon atoms, while $|\psi|^2$ of $C$ and $D$ states are completely
zero in this region.} \label{fig2}
\end{figure}

The gapless chiral edge states can be clearly seen from the energy
spectrum of graphene ribbons. Fig.\ref{fig2}.(a) plots the band
structure of graphene ribbons, with zigzag edges and 800 atoms
across the width. The parameters used here are $t_{so}=0.2$ and
$\lambda=0.18$. One can easily distinguish the gapless edge states
from the bulk states. For a given Fermi level in the gap, there
exists four different edge states labeled as $A$, $B$, $C$, and $D$.
From
$\bm{v}(\bm{k})=\frac{1}{\hbar}\frac{\partial{E(\bm{k})}}{\partial{\bm{k}}}$,
one can find that states $A$ and $B$ ($C$ and $D$) propagate along
the same $-x$ ($+x$) direction. Panel (b) plots the wave function
distributions $|\psi|^2$ of the four states across the width (only
part of the ribbons with one zigzag edge are plotted). One can
observe that the wave functions of the $A$ and $B$ states are
localized at the left boundary, with noticeable values on about 30
carbon atoms, whereas the wave functions of the $C$ and $D$ states
are essentially zero in this region. The contrary occurs on the
other boundary (not shown in the figure). This is topologically
distinct from the helical edge states of the quantum spin Hall
effect, where opposite spins propagate in opposite directions along
the same boundary.

\begin{figure}
\includegraphics[width=6.5cm,totalheight=6cm,angle=0]{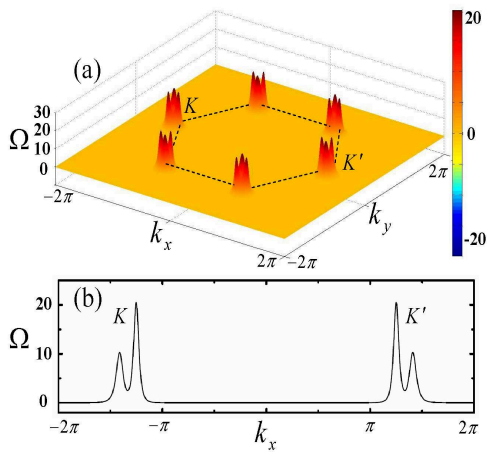}
\caption{(Color online) (a) Berry curvature distribution $\Omega$
(in units of $e^2/h$) of the valence bands in the momentum space.
The first Brillouin zone is outlined by the dashed lines, and two
inequivalent valleys are labeled as $K$ and $K'$. (b) The profile of
Berry curvature distribution along $k_y=0$. The parameters used here are
$t_{so}=0.1$ and $\lambda=0.18$.} \label{fig3}
\end{figure}

The emergence of chiral edge states in the bulk gap is intimately
related to the topological property of the bulk Bloch states in the
valence bands. This is characterized by the quantized charge Hall
conductance: $\sigma_{yx}=\mathcal{C}~{e^2}/{h} $, where
$\mathcal{C}$ is an integer known as the Chern
number~\cite{Thouless,Kohmoto} and can be calculated from:
\begin{eqnarray}
\mathcal{C}=\frac{1}{2\pi}\sum_n\int_{\text{BZ}}d^2k \Omega_n
\label{Chern}
\end{eqnarray}
where $\Omega_n$ is the momentum-space Berry curvature for the
$n$-th band~\cite{MChang,Thouless,yao1}
\begin{eqnarray}
\Omega_n(\bm{k})=-{\sum_{n^{\prime} \neq n}} {\frac{2 {\rm {Im}}
\langle \psi_{n \bm{k}}|v_x|\psi_{n^\prime \bm{k}} \rangle \langle
\psi_{n^\prime \bm{k}}|v_y|\psi_{n \bm{k}} \rangle }
{(\omega_{n^\prime}-\omega_{n})^2}} \label{berry}
\end{eqnarray}
The summation is over all occupied bands below the bulk gap,
$\omega_n\equiv E_n/ \hbar$, and $v_{x(y)}$ is the velocity
operator. The absolute value of $\mathcal{C}$ corresponds to the
number of gapless chiral edge states along an edge of the 2D
system~\cite{Chern_Edge}. In Fig.\ref{fig3}.(a), we plot the Berry
curvature distribution $\Omega$ for the valence bands in the
momentum space. Panel (b) shows the profile of the Berry curvature
along the $k_y=0$ intersection. We observe that the Berry curvature
is peaked at the corners of the first Brillouin zone. It has the
same sign at the inequivalent $K$ and $K'$ points, because the
honey-comb lattice preserves the two dimensional inversion symmetry
for which $\Omega(\bm{k})=\Omega(-\bm{k})$. This is in contrast with
the valley Hall effect, found in graphene system with AB sublattice
symmetry breaking, where the Berry curvature at $K$ and $K'$ points
has opposite signs~\cite{XiaoDi}. In the semiclassical picture, the
curvature field acts on the carriers as a magnetic field in the
momentum space, and deflects the carriers in transverse direction
under a driving force. We obtain the Chern number from the curvature
integration and find that $\mathcal{C}=2$, in agreement with the
number of chiral edge states from tight-binding calculations.

\begin{figure}
\includegraphics[width=7.5cm,totalheight=3.5cm,angle=0]{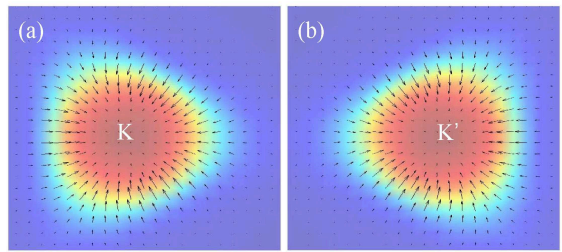}
\caption{(Color online) (a)-(b) Top view of the spin configurations
of the higher valence band in the $k_x$-$k_y$ plane around $K$ and
$K^\prime$ points. Blue/red regimes represent the spin-up/down
states. The arrows represent the spin direction at each
point.} \label{fig4}
\end{figure}

There is a simple explanation of the Chern number value
$\mathcal{C}=2$. As shown in Fig.\ref{fig1}.(d), there are two
valence bands: the lower one is almost completely spin-down, while
the higher one involves a strong variation of the spin direction in
the momentum space. By plotting the spin configuration in
$k_x$-$k_y$ plane~(see Fig.\ref{fig4}), we find that the spin
textures around $K$ and $K'$ form two separate Skyrmions of the same
sign~\cite{Skyrmion}: spins point down in the central region (shown
in red color), but point up in the outside region(shown in blue
color). Since each Skyrmion contributes to a Chern number
$\mathcal{C}=1$, Skyrmions at $K$ and $K'$ points give rise to a
total Chern number $\mathcal{C}=2$ in our model.

\begin{figure}
\includegraphics[width=8cm,totalheight=9.5cm,angle=0]{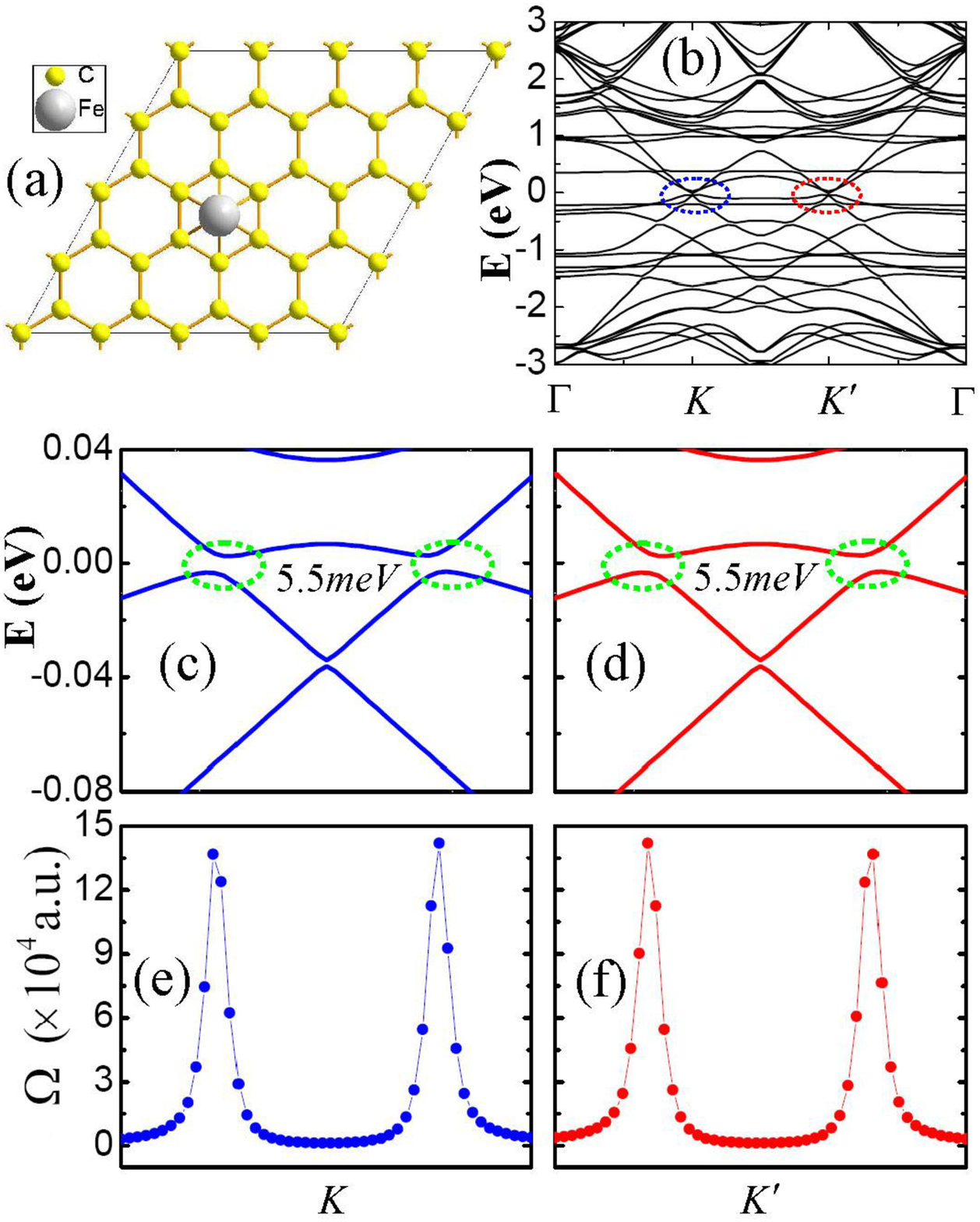}
\caption{(Color online) (a) Typical structure geometry of a Fe atom
on top of the hollow position of a $4\times4$ supercell of graphene.
(b) Bulk band structure with SOC included along the high symmetry
lines. The Fermi level is exactly located in the gap. (c)-(d)
Zooming in of the band structure at $K$ and $K'$ points. A gap
$\sim5.5~\textrm{meV}$ circled by the green dotted curves is opened.
(e)-(f) Berry curvature distribution $\Omega$ of the valence
bands near $K$ and $K'$ points. Note: the Berry curvature at $K$ and
$K'$ shares the same sign thus giving a non-zero Chern number
$\mathcal{C}$.} \label{fig5}
\end{figure}

So far, we have predicted the occurrence of the quantum anomalous
Hall effect for a model Hamiltonian with Rashba SOC and exchange
field uniformly distributed on the honey-comb lattice. In the
following, we consider another model in a similar spirit but with a
non-uniform distribution: a graphene sheet with magnetized
transition metal atoms adsorbed on one side (i.e., top). Here we
study a concrete case of Fe adsorbed graphene using first principles
method, and results with other kinds of adsorbates will be published
in a separate paper~\cite{yao2}. In our calculations, all atoms are
allowed to relax along the normal direction of the graphene sheet,
but we have kept the experimental value of the lattice constant for
graphene in the $x$-$y$ plane. Fig.\ref{fig5}.(a) shows a unit cell
of a superlattice, with one Fe atom adsorbed on top of each
$4\times4$ supercell of the graphene sheet. The calculations were
performed using the projected-augmented-wave (PAW) method~\cite{PWA}
as implemented in the VASP package~\cite{VASP}, and the GGA exchange
correlation potential~\cite{GGA} was used.

We find that the most stable adsorption positions are on top of the
hollow centers (with an equilibrium separation of $1.56~{\AA}$),
which is consistent with the previous study without
SOC~\cite{Cohen,Kang}. Fig.\ref{fig5}.(b) shows the bulk band
structure with SOC of the iron taking into account. The magnetic
moment of iron is found to be about 2$\mu_B$ in our calculations.
This is consistent with the fact that there are totally eight
electrons in the 3-\emph{d} orbitals of the Fe atom after two
4-\emph{s} electrons being transferred there~\cite{Cohen}.
Therefore, no electron transfer occurs to the carbon atoms, thus the
Fermi level is exactly located in the gap. Figs.\ref{fig5}.(c) and
(d) zoom in the band structure around $K$ and $K'$ points circled in
panel (b). One can observe that a bulk gap around $5.5~\textrm{meV}$
is opened at both $K$ and $K'$ points. After obtaining the Bloch
functions from the self-consistent potentials, the corresponding
Berry curvature $\Omega(\bm{k})$ near $K$ and $K'$ points are
calculated from the Kubo-formula~\cite{Thouless,yao1}: $
\Omega(\bm{k})=-{\sum_{n}}{f_{n}\Omega_n(\bm{k})}\label{berry}, $
where $f_n$ is the equilibrium Fermi-Dirac distribution function.
Figs.\ref{fig5}.(e) and (f) plot the Berry curvature distribution
$\Omega$ summing over all the valence bands. Two spikes near $K$ and
$K'$ are clearly seen, and a Chern number of $\mathcal{C}=2$ for the
valence bands is obtained by integrating over the whole Brillouin
zone. This agrees with the result of our tight-binding model,
despite the fact that the exchange and spin-orbit effects produced
by the Fe atoms are clearly not uniform over the lattices.

In summary, we find that a non-trivial bulk gap in graphene can be
produced in the presence of both Rashba SOC and exchange field,
where we predict a quantum anomalous Hall conductance of
$\sigma_{yx}$ quantized as $2e^2/h$. This is followed up with a more
concrete example of graphene sheet with Fe atoms adsorbed on top.
Our first principles calculations show a bulk gap as large as
$\sim5.5~\textrm{meV}$ can be opened at the Dirac points , producing
the same topological effect.

Note added: After the completion of the bulk of this work (see, the
brief announcement in Ref.~\cite{abstract}), we noticed a similar
work~\cite{QAHE_Bilayer} demonstrating the quantum anomalous Hall
effect in the bilayer graphene.

Z.Q. was supported by NSF~(DMR0906025) and Welch
Foundation~(F-1255). Q.N. was supported by DOE~(DE-FG02-02ER45958,
Division of Materials Science and Engineering) and Texas Advanced
Research Program. Y.Y. was supported by NSF of China~(10674163,
10974231) and the MOST Project of
China~(2006CB921300,~2007CB925000). J.W. was supported by RGC
grant~(HKU7054/09P) from the government of SAR of Hong Kong and
LuXin Energy Group. The Texas Advanced Computing Center and Computer
Center of the University of Hong Kong are gratefully acknowledged
for computing assistance.

$^*$ygyao@aphy.iphy.ac.cn

\end{document}